\documentclass[twocolumn,showpacs,amsmath,amssymb,prd,floatfix]{revtex4}
\usepackage{graphicx}
\usepackage{dcolumn}
\usepackage{bm}

\def\beq{\begin{equation}}
\def\enq{\end{equation}}
\def\ba{\begin{eqnarray}}
\def\ea{\end{eqnarray}}
\def\<{<\!\!}
\def\>{\!\!>}
\def\ra{\rightarrow}

\begin{document}
\input{epsf}

\title{Astrophysical tau neutrino detection
in kilometer-scale Cherenkov detectors via muonic tau decay}

\author{T. DeYoung$^1$, S. Razzaque,$^{2,1}$, and D. F. Cowen,$^{1,2}$}

\affiliation{$^1$Department of Physics,
Pennsylvania State University, University Park, PA 16802}
\affiliation{$^2$Department of Astronomy \& Astrophysics,
Pennsylvania State University, University Park, PA 16802}

\begin{abstract}
Kilometer-scale deep under-ice or -water Cherenkov
neutrino detectors may detect muon and electron neutrinos from 
astrophysical sources at energies of a TeV and above.  Tau neutrinos 
are also expected from these sources due to neutrino flavor 
oscillations in vacuum, and tau neutrinos are free of atmospheric 
background at a much lower energy than muon and electron neutrinos.  
Identification of tau neutrinos is expected to be possible above 
the PeV energy range through the ``double bang'' and ``lollipop'' 
signatures.  We discuss another signature of tau in the 
PeV--EeV range, arising from the decay of tau leptons inside the detector to 
much brighter muons.
\end{abstract}

\date{\today}
\maketitle

\section{Introduction}

Kilometer-scale Cherenkov neutrino detectors now planned or under
construction, such as IceCube
\cite{Ahrens:2003ix} at the South Pole and KM3NeT \cite{km3net}
in the Mediterranean, are poised to detect high energy neutrinos 
from astrophysical sources such as gamma-ray bursts (GRBs) and active 
galactic nuclei (AGN).  Ultrahigh energy (UHE) cosmic rays with 
energies exceeding $\sim\!\! 10^{20}$ eV are thought to originate 
from these extraordinary sources of the highest energy $\gamma$-rays 
observed. High energy neutrinos should also be produced in these 
sources as the result of photomeson ($p\gamma$) and/or proton-proton 
($pp$) interactions of shock accelerated protons with ambient 
radiation fields and/or plasma material.

After propagating virtually unimpeded through the universe, high
energy neutrinos detected by experiments on Earth may reveal the
physical and astrophysical conditions of their sources at energies and
distances unmatched by other methods.  In addition to measuring the 
energy and direction of astrophysical neutrinos, future neutrino 
telescopes will be able to distinguish between the three known flavors 
of neutrinos (and anti-neutrinos), namely electron, muon and tau 
neutrinos ($\nu_e$, $\nu_{\mu}$ and $\nu_{\tau}$), by looking at the
signature(s) of their interactions in the detection media.

Tau neutrinos are not produced in appreciable numbers in astrophysical 
sources, but will appear in numbers comparable to $\nu_e$ and $\nu_\mu$
as a consequence of flavor oscillation between their sources and Earth.  
Astrophysical neutrino sources generically are believed to produce neutrinos
through $\pi^\pm$ (and $K^\pm$) decay, leading to a neutrino flavor ratio at 
production of $\nu_{e} : \nu_{\mu}: \nu_{\tau} = 1:2:0$.  However, neutrino
flavor eigenstates $\nu_{\alpha}$ ($\alpha=e,\mu,\tau$) and mass
eigenstates $\nu_{j}$ ($j=1,2,3$) are mixed by a unitary matrix
$U$ defined as $\nu_{\alpha} = \sum_{j} U^{*}_{\alpha j} \nu_{j}$.
The oscillation probability is given by ${\cal P}_{\nu_{\alpha}
\ra \nu_{\beta}} = |\sum_{j} U_{\beta j} e^{-i\varphi}
U^{*}_{\alpha j}|^2$, where 
	\[ 
	\varphi = 6.3 \cdot 10^9 
	\left(\frac{\delta m^2}{8 \cdot 10^{-5} \; \rm{eV^2}}\right)
	\left(\frac{D}{\rm{kpc}}\right) \left(\frac{\rm{TeV}}{E_{\nu}}\right)
	\]
is the phase of the slower (solar) neutrino oscillation and $D$ is the 
distance traveled by the neutrinos before being detected.  The oscillation
probability reduces to ${\cal P}_{\nu_{\alpha} \ra \nu_{\beta}} 
\approx \sum_{j} \left|U_{\beta j}\right|^2 \cdot \left|U_{\alpha 
j}\right|^2$ for $\varphi \gg 1$, which holds for essentially all 
astrophysical sources.  We use the
standard expression for $U_{\alpha,j}$ from Ref.~\cite{pdg04} with
solar mixing angle $\theta_{\odot} \equiv \theta_{12} =
32.5^{\circ}$ and the atmospheric mixing angle $\theta_{\rm atm}
\equiv \theta_{23} = 45^{\circ}$, following the results from the
SNO \cite{Ahmed:2003kj} and K2K \cite{Ahn:2004te} experiments, respectively. 
The unknown mixing 
angle $\theta_{13}$ and the CP violating phase may be assumed to be zero
given the current upper bounds from reactor experiments. The
resulting astrophysical neutrino flux ratio on Earth is thus generally 
expected to be $\nu_{e} : \nu_{\mu} : \nu_{\tau} = 1:1:1$, 
although this may be modified somewhat by effects such as neutron
decay, two-photon annihilation to muon pairs, or muon synchrotron 
cooling in the source environment \cite{Stanev:2000fb, Razzaque:2005ds, 
Rachen:1998fd, Kashti:2005qa}.
  
Tau neutrinos are particularly interesting because
local backgrounds to astrophysical $\nu_\tau$ signals are low.  The possible 
backgrounds are from UHE cosmic rays interacting in Earth's atmosphere 
and producing short-lived charmed mesons which decay as 
$D_s \ra \tau \nu_{\tau}$, known as the ``prompt'' neutrino flux
\cite{Gondolo:1995fq, Pasquali:1998xf}, and from conventional atmospheric 
$\nu_e$ or $\nu_\mu$ produced in these cosmic ray air showers oscillating
to $\nu_\tau$ as they traverse the Earth before being detected.

The prompt atmospheric $\nu_{\tau}$ flux is not precisely known
due to uncertainties in the extrapolation of the parton distribution
functions to low $x$ and in the composition of the high energy
cosmic rays.  However, the different models lead to expected 
rates of $10^{-3}$ to $10^{-2}$ prompt tau events above 1 PeV 
per year in a km$^3$ neutrino telescope 
\cite{Pasquali:1998xf, Athar:2001jw, Dutta:2000jv, Martin:2003us,
Beacom:2004jb}.   

As for oscillations to $\nu_\tau$, the maximum propagation distance
$L\approx 10^4$ km, the Earth's 
diameter, corresponds to a $\nu_{\mu} \ra \nu_{\tau}$ oscillation 
probability ${\cal P} \approx 10^{-3}(E_{\nu}/{\rm TeV})^{-2}$
(neglecting matter effects, which would reduce the oscillations), 
so oscillations will produce a negligible flux of $\nu_\tau$ for 
$E_\nu$ at the PeV scale, even in comparison to the low prompt 
fluxes.   The only remaining source of PeV-scale tau neutrinos is
extraterrestrial.

Tau neutrinos are also interesting because of the phenomenon of $\nu_\tau$ 
regeneration \cite{Halzen:1998be}.  The high energy neutrino interaction cross
section rises approximately linearly with energy, and the mean free path
through the Earth becomes shorter than the Earth's diameter for 
$E_\nu \gtrsim 100$ TeV.  The Earth is thus opaque to $\nu_e$ and $\nu_\mu$ 
at PeV energies and above, and only horizontal and downgoing neutrinos can be
observed.  For $\nu_\tau$, however, the $\tau^\pm$ produced in a charged 
current (CC) neutrino interaction will usually decay back to $\nu_\tau$ before
losing significant amounts of energy, effectively regenerating the $\nu_\tau$
beam and leaving an upgoing $\nu_\tau$ flux up to the PeV scale.

The best-known signature for detecting $\nu_{\tau}$ in a water or ice
Cherenkov detector is called the 
``double bang'' \cite{Learned:1994wg}.  In these events, a 
CC neutrino-nucleon interaction $\nu_{\tau}N \ra 
\tau X$ produces a hadronic shower (denoted $X$), with the subsequent 
decay of the $\tau$ lepton producing a second shower, connected to 
the first by the $\tau$ lepton track.  The second shower may 
also be hadronic, or it may be electromagnetic in the case of 
$\tau \ra e \nu_\tau \bar{\nu_e}$.  The $\tau$ produced in the CC 
interaction has energy $\left\langle E_\tau \right\rangle \approx 
0.75 E_\nu$ \cite{Gandhi:1995tf}, and the two showers are separated 
by the tau decay length 
$l_\tau = \gamma c t_\tau \sim 50 \, (E_{\tau}/{\rm PeV}) 
\, {\rm m}$ (neglecting energy losses along the track).  Due to the 
short $\tau$ lifetime and wide spacing of the detection 
elements in kilometer-scale neutrino telescopes, this signature is only 
expected to be detectable for $\nu_{\tau}$ with energy $E_{\nu} 
\gtrsim$ PeV\@.  Above $\sim 20$ PeV, the typical decay length exceeds 
1 km, so both showers 
usually will not be contained in a kilometer-scale detector; the 
resulting signature of a tau track and one shower is known as a 
``lollipop'' \cite{Beacom:2003nh}.

In this paper, we point out another distinctive signature of 
extremely high energy (EHE, PeV--EeV) tau leptons, produced by the 
muonic decay of a $\tau$ inside 
the instrumented detection volume.  Although the muon has lower energy
than the parent tau lepton, it will emit more light than the tau.  The 
lepton track will thus appear to suddenly increase in brightness by an 
amount which should be detectable in a neutrino telescope.  

The energy
range over which this signature is observable is constrained at the lower
end by the requirement that a reasonably long tau lepton track be observed
prior to the tau decay.  At the higher end, the rising rate of tau
photonuclear energy loss causes the brightness of the tau to 
approach that of the daughter muon above EeV energies.  It should be noted
that these limits apply to the energy of the $\tau$ lepton in the detector; 
events from higher energy $\nu_\tau$ could be observed if the initial neutrino 
interaction vertex is some distance from the detector so that the $\tau$ 
lepton loses energy in stochastic interactions before decaying within the 
detector.

\section{Signature of Muonic Decay}

Identification of tau events through muonic decay, $\tau \, \ra 
\! \mu \nu_{\tau} \nu_{\mu}$, requires the decay to occur within the 
detector so that the increase in 
brightness will be observed.  The branching ratio for 
this decay channel is measured to be $\Gamma_{\mu} = 17.36\%$ 
\cite{pdg04}, so only a fraction of tau leptons will manifest themselves via
this signature.  However, at energies $\gtrsim 20$ PeV, 
the tau track length $L_\tau$ becomes longer than the geometric
scale of the detector and double bangs are no longer visible.  
Only the lollipop and muonic decay signatures can be used to identify 
taus in this regime, even though only a fraction 
$\sim (1 \:{\rm km}) / L_\tau$ of the 
taus can be tagged by these methods; lollipops are generally
considered to be identifiable only when the final, not the initial, 
bang is observed (see Section \ref{sec:discussion}).

Tau leptons are produced in neutrino $V-A$ interactions, which 
at the energies of interest produce polarized taus.  The spectrum
of muon energies from the decay of polarized $\tau$ is 
$dn/dx = \frac{4}{3}(1-x^3)$, where $x = E_\mu / E_\tau$ 
\cite{Gaisser92, Lipari:1993hd, Dutta:2000jv}.  The expected muon 
energy is thus
\[
	\left\langle E_\mu \right\rangle = 0.4 \: E_\tau \: . 
\]
Depolarization via $\tau$ interactions prior to decay should be
small, and would reduce the muon energy only slightly, to 
$0.35 E_\tau$ in the limit of complete depolarization 
\cite{Gaisser92}.

Although the muon has less than half the energy of the tau, it 
appears brighter because the muon loses energy more rapidly than 
the tau, as discussed below.

\subsection{Energy Loss Rates}

The average energy loss of heavy leptons per
unit distance traveled in matter (in g/cm$^{2}$) is often 
approximated as $- \left\langle dE/dX \right\rangle \approx a+bE$.  
The constant part, due to ionization losses,
may be calculated using Bethe-Bloch formula, while the radiative part
approximately proportional to $E$ is due to a combination of 
stochastic $e^+e^-$ pair production, bremsstrahlung and photonuclear 
effects.  In the EHE regime, the ionization loss term $a$ is negligible 
compared to the stochastic losses.  The radiative energy loss parameter 
$b = b(E)$ varies slowly with energy for extremely high energy leptons, 
primarily due to an increase in the photonuclear energy loss rate $b_{pn}$ at 
very high energies.  Although the radiative energy losses are in fact
due to a series of discrete stochastic events, at high energies 
these interactions occur frequently enough that they can be 
considered quasi-continuous, increasing the overall
brightness of the lepton track.

While $e^+e^-$ pair production and bremsstrahlung are the dominant 
energy loss channels for $\mu$ above the TeV scale, for $\tau$
bremsstrahlung is negligible and photonuclear effects dominate
at EHE.  Tau 
photonuclear losses are comparable to electromagnetic 
losses all the way down to the ionization-dominated region below 
$E_\tau \sim 20$ TeV \cite{Dutta:2000hh, Bugaev:2003sw}. 

\begin{figure}[htbp]
	\centering
		\includegraphics[width=1.0\columnwidth]{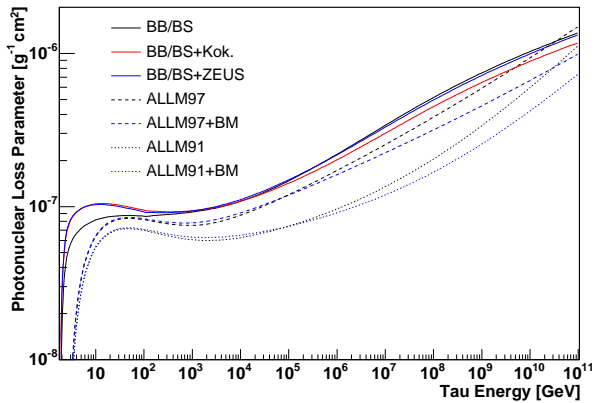}
		\caption{Photonuclear energy loss per unit energy ($b_{pn}$) for $\tau$ 
			leptons in ice, according to various models.  BB/BS refers to 
			\cite{Bezrukov:1981ci}, plus the hard component of \cite{Bugaev:2002gy}.  
			Kok and ZEUS include the photon-nucleon cross sections of 
			\cite{Kokoulin:1999bn} and \cite{Derrick:1994dt}, respectively, instead of 
			that from \cite{Bezrukov:1981ci}.  ALLM91 and ALLM97 refer to 
			\cite{Abramowicz:1991xz} and \cite{Abramowicz:1997ms}, with BM following 
			\cite{Abramowicz:1991xz, Abramowicz:1997ms} but using the nuclear structure
			function from \cite{Butkevich:2001aw}.}
	\label{fig:TauPhotonucIce}
\end{figure}

Photonuclear energy losses by EHE leptons (and to a lesser 
extent bremsstrahlung losses) are not precisely known.  Measurements of 
photonuclear cross sections from 
colliders must be extrapolated to very small $x$, and there are several 
models available in the literature \cite{Bezrukov:1981ci, Bugaev:2002gy, 
Bugaev:2003sw, Kokoulin:1999bn, Derrick:1994dt, Abramowicz:1991xz, 
Abramowicz:1997ms}.  Different models of the nuclear structure function are 
also available \cite{Abramowicz:1991xz, Abramowicz:1997ms, Butkevich:2001aw}.
The predicted loss rates for $\tau$ leptons in ice are shown in 
Fig.~\ref{fig:TauPhotonucIce}.  Numerical values for energy loss rates given 
in this paper were evaluated using the MMC software package 
\cite{Chirkin:2004hz, Chirkin:2005hu}, using ice as the default detection 
medium.  

Recent calculations of muon 
bremsstrahlung \cite{Kelner:1995hu, Kelner:1997cy, Andreev:1994pr, 
Bezrukov:1981ca} are in close agreement with each other (although 
they are about 5\% higher than older calculations 
\cite{PetrukinShestakov}), but disagree by as much as 20\% for EHE taus.  
Because bremsstrahlung is suppressed by $1/m_l^2$, however, the 
differences between models have very little effect on the overall brightness 
of $\tau$ tracks.

\subsection{Photonuclear Interactions}
\label{sec:photonuc}

In addition to the uncertainty in photonuclear energy loss rates
at high energies noted above, there are several differences between
photonuclear and electromagnetic interactions that must be considered.

Both bremsstrahlung and pair production produce electromagnetic
showers in the Cherenkov medium, with $\gamma$'s converting to 
$e^+e^-$ pairs and the $e^+$ and $e^-$ in turn radiating more 
$\gamma$'s.  Photonuclear interactions, on the other hand,
disrupt the nucleon involved and produce showers of hadrons
which go on to interact with other nucleons in the medium.  

The light produced in a shower comes from the Cherenkov radiation
of the many secondary particles involved in the shower.  The
total light yield is proportional to the integrated track length 
of the relativistic particles.  The yield per unit shower energy 
is lower in hadronic showers because heavy particles have a higher 
threshold for Cherenkov radiation, energy is lost to 
the binding energies of the hadrons involved, and invisible slow neutrons 
are produced \cite{Gabriel:1993ai}.  

The ratio of the light yield per unit energy in hadronic showers 
to that in electromagnetic showers depends on energy, as well.  Neutral $\pi$ 
mesons produced in the shower will decay to $\gamma \gamma$ and produce 
electromagnetic subshowers, but very few hadronic particles are formed in
electromagnetic showers.  The production of $\pi^0$ is thus a ``one-way 
street'' \cite{Gabriel:1993ai} carrying energy out of
the hadronic sector and into the electromagnetic.  Roughly 30\% of the energy 
in hadrons will go to $\pi^0$ in each generation of the
shower, so high energy hadronic showers with more generations will be more
like electromagnetic showers.  For small ($\sim 10$ 
GeV) showers the ratio of light yields is about 65\%, rising to 
about 85\% for a 10 TeV shower \cite{Wiebusch:1995bw, Kowalski:2004qc} and
asymptotically approaching 100\%.  In the calculations presented in 
Sect.~\ref{sec:detectability} we have assumed an average value of 75\% for
the energy deposited in hadronic photonuclear interactions along the $\tau$
(and $\mu$) track.

Another important effect to consider is the energy spectrum of
individual stochastic interactions.  Relatively large fractions
$v$ of the total lepton energy can be deposited in a single 
photonuclear interaction.  In the analysis of Ref.~\cite{Bugaev:2002gy} in 
terms of a hard (perturbative) and a 
soft (non-perturbative) component, the energy spectrum 
$v \:\, d\sigma / dv$ of the individual photonuclear 
interactions comprising the perturbative component peaks at 
$v \sim 10^{-2}$ for $\tau$ leptons.  Since the radiative energy 
loss parameter $b \lesssim 10^{-6}$ g$^{-1}$ cm$^2$, such 
interactions will be separated by several kilometers.  

Most often, such perturbative interactions will not occur along the
contained track segment even in a kilometer-scale neutrino 
telescope.  In this case, the perturbative component of the
photonuclear term (which in the analysis of Ref.~\cite{Bugaev:2002gy} 
dominates above $\sim 10$ PeV) should be ignored in estimating 
the brightness of the tau track, and only the non-perturbative component 
should be taken into account.  If a perturbative interaction \textit{does} 
occur within the detector, it will appear as a distinct cascade along the
track, and it should still be possible to measure the 
``baseline'' track brightness away from the shower (although
the bright shower could confuse event analysis algorithms and 
thus lower the overall efficiency for detecting such events).  

As the tau energy increases over the EHE regime, however, 
$v \:\, d\sigma / dv$ shifts to lower $v$ and the total photonuclear energy 
loss parameter $b_{pn}$ increases.  Perturbative photonuclear interactions 
thus become somewhat softer and much more common, with separations of a few 
hundred meters.  In this regime, the quasi-continuous approximation may be
appropriate for the perturbative component as well as the non-perturbative
part, although the details will depend on the 
particular detector and event reconstruction algorithm under 
consideration.  In Sect.~\ref{sec:detectability} we present results for the 
two limiting cases, both of which should produce a detecatable signal: first 
assuming the perturbative photonuclear interactions of the 
$\tau$ are very rare, so that only the non-perturbative component of the 
photonuclear interactions contributes to the brightness of the $\tau$ track; 
and then assuming the perturbative interactions are also quasi-continuous, 
increasing the average brightness of the $\tau$ track and thus decreasing the 
ratio of the brightness of the tau to that of the muon.

\subsection{Radiative Decays}

Tau decays to $\mu$ are sometimes accompanied by initial-state or final-state 
radiation.  This radiation follows a $1/k$ spectrum at low energies (below the
$\tau$ mass scale), and is measured to occur with a branching fraction of 
0.36\% with a threshold of $E_\gamma > 10$ MeV in the $\tau$ rest frame 
\cite{pdg04}.  These photons will be boosted by a factor of up to 
$\gamma \,(1+\beta)$, depending on the emission angle of the photon in the
$\tau$ rest frame,
so that for $E_\tau = 1$ PeV a photon with 10 MeV in the 
rest frame will have at most $\sim \! 20$ TeV in the detector frame.  Neutrino
telescopes generally have an energy threshold of several TeV to a few tens of 
TeV for reconstruction of cascades (e.g. 
$\nu_e$ CC events), so 20 TeV is a reasonable benchmark
for the energy at which a shower from initial-state or final-state radiation
might be noticeable along a lepton track.  For $\tau$ tracks at the PeV scale,
we therefore expect a sizeable shower at the decay vertex in 
2\% or fewer of observed tau decays.  Because of the higher boost 
factor, noticeable showers will be about twice as common for EeV $\tau$ 
decays.  It is not clear that these showers would be distinguishable from the 
background of stochastic energy losses, but it might be helpful to treat 
sizeable showers along an observed track as candidate positions for tau decay 
vertices.

\section{Detectability}
\label{sec:detectability}

\begin{figure}[htbp]
	\centering
		\includegraphics[width=1.0\columnwidth]{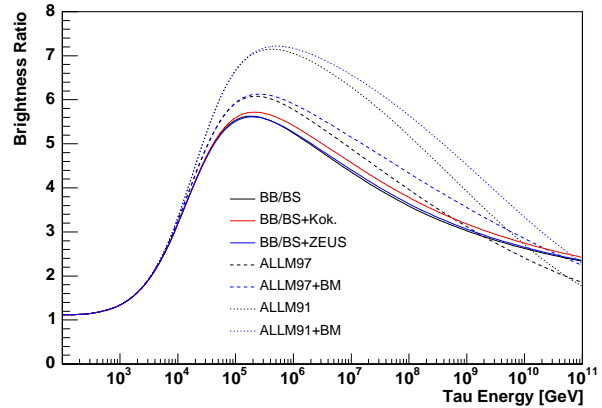}
		\caption{The magnitude of the increase in brightness as a $\tau$ decays to
			$\mu$, as a function of the $\tau$ energy.  The $\mu$ is assumed to take
			$1-\left\langle y \right\rangle = 0.4$ of the $\tau$ energy.  The various
			lines correspond to different models for photonuclear energy loss, as in 
			Fig.~\ref{fig:TauPhotonucIce}.  In the region of interest from 1 PeV to
			1 EeV, the brightness steps up by a factor of between 3 and 7, depending
			on the model and on the energy of the $\tau$.}
	\label{fig:BrightnessRatio}
\end{figure}

An EHE $\tau$ decaying to $\mu$ will appear as a track which 
suddenly increases in brightness.  The magnitude of the increase, taking into 
account the most probable fraction of the tau energy carried by the muon, the 
average energy loss rates of the two leptons, and the relative light yields 
of hadronic and electromagnetic showers, is shown in 
Fig.~\ref{fig:BrightnessRatio}.  Losses to ionization, pair production and 
bremsstrahlung are also included; the bremsstrahlung model of 
\cite{Kelner:1997cy} is used, but the results are nearly independent of the
choice of bremsstrahlung model.

Although the IceCube collaboration has not published any estimate of 
track energy resolution for the IceCube detector, a resolution of 
$\sigma(\log_{10} E_\mu) \simeq 0.3$, corresponding to a factor of 2 in 
$E_\mu$, was claimed for AMANDA-II \cite{Ahrens:2003fg}.  (In the 
radiative-dominated regime, the
brightness scales approximately linearly with muon energy, so the resolution
in brightness should be the same as the energy resolution.) One would expect 
IceCube to do at least this well, given the larger detector volume and better 
optical module electronics.  

The ANTARES collaboration expects a track energy resolution of 
$\sigma(\log_{10} E_\mu) \simeq$ 0.3--0.4 for PeV muons \cite{Carr:2004wy}
(resolution at higher energies was not given).  A review of the literature 
did not produce published track energy resolutions for the Baikal, NESTOR, or
NEMO detectors.  These smaller detectors (like AMANDA) do not
have the effective volume to yield significant event rates at and above the
PeV scale in any case, but the energy resolution for km$^3$ detectors in
water should be comparable to, or better than, the ANTARES resolution.

It appears that, for both ice and water km$^3$ neutrino telescopes, the 
expected track energy resolution should be sufficient to distinguish the
brightness of the initial $\tau$ track from that of the final $\mu$ track.  We
note that the energy resolutions given above refer to measuring a single
energy for a through-going track, rather than trying to make separate energy 
measurements for different segments of a track.  However, in a km$^3$
detector, the observed tracks will be several times longer than those visible
in smaller instruments such as AMANDA or ANTARES, so enough information
should be recorded for measurements of comparable accuracy.

At the lower end of the EHE scale considered here (a few PeV), the $\tau$
track will only be visible if the neutrino interaction vertex is contained
within the detector; even then, the $\tau$ track will be quite short.  
Reconstruction of such events will be further complicated by the hadronic
shower produced at the $\nu N$ vertex, which will have typically an energy of
$0.25 E_\nu$.  The ability to measure the energy of a short track segment 
within such a shower must be determined with a detailed and detector-specific
Monte Carlo study, but it seems likely that a $\tau$ track with a length of at
least 200 m will be required.  This would correspond to a neutrino energy 
threshold of around 5 PeV for detection of $\tau$ using this signature,
comparable to but slightly higher than the threshold for the double bang
signature.  The advantage of the muonic signature is that it should be 
remain detectable to very high energies, at least up to the EeV scale.

\begin{figure}[tbp]
	\centering
		\includegraphics[width=1.0\columnwidth]{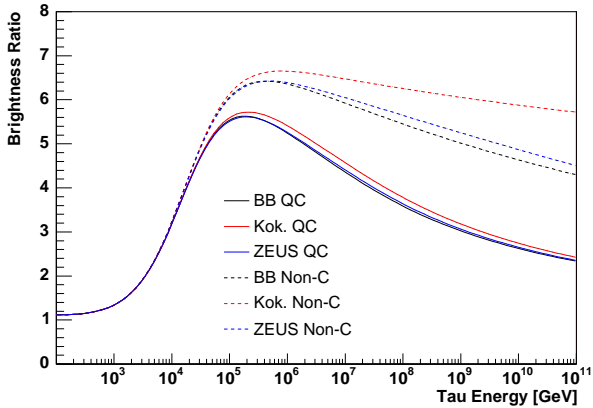}
		\caption{The increase in brightness as $\tau$ decays to $\mu$, under the
			alternative assumptions that: (QC) the hard component of the photonuclear 
			interaction is quasi-continuous, or (Non-C) that it is rare enough that it 
			does not affect the measurement of the track brightness.  The ability to 
			resolve such showers and differentiate them from the track will depend on 
			the neutrino detector under consideration.}
	\label{fig:BrightQuasiCont}
\end{figure}

As discussed in Sect.~\ref{sec:photonuc}, the curves in 
Fig.~\ref{fig:BrightnessRatio} assume that the light emission from 
photonuclear interactions is quasi-continuous, which may not be a valid
assumption.  Figure~\ref{fig:BrightQuasiCont} shows, for the model of 
\cite{Bugaev:2002gy}, the effect of assuming to the contrary that hard 
photonuclear interactions are sufficiently rare that they do
not contribute to the measured brightness of the underlying track.  This 
assumption would hold if no such interactions occurred within the detector
volume.  Alternatively, an analysis searching for muonic tau decays could 
attempt to identify bright showers along the observed track, and measure the
``baseline'' track brightness away from such showers; because very hard 
stochastic interactions are more common for $\tau$ than for $\mu$, such an 
approach should heighten the contrast between $\mu$ and $\tau$ tracks.  To our
knowledge, no study of the ability of any neutrino telescope to resolve
showers along a track in this manner has yet been published.

\section{Discussion}
\label{sec:discussion}

The detection of tau neutrinos offers an excellent method for detecting
astrophysical neutrinos, since there is essentially no atmospheric tau 
neutrino background.  However, the classic double bang signature of tau
neutrinos is only expected to be detectable over a single decade of 
energy, from a few PeV to perhaps 20 PeV.  The possibility of tagging tau
events at higher energies by detecting only the final bang, known as a 
lollipop event, has already been pointed out.  In this paper, we have shown 
that it should also be possible to tag taus which decay to muons.

We note that this signature of $\tau$ decaying to $\mu$ was considered in
Ref.~\cite{Bugaev:2003sw}, which concluded that the increase in brightness
would not be detectable in a neutrino telescope.  Two factors lead us to the
opposite conclusion: the typical fraction of energy transferred to the $\mu$
is somewhat higher than they assumed, and the fact that a $\tau$ loses energy
primarily through photonuclear interactions reduces the \emph{brightness} of 
the $\tau$ track relative to that of a $\mu$ losing energy at the same rate.

We also note the observation made in Ref.~\cite{Bugaev:2003sw} that the double
bang and lollipop signatures are not completely free of experimental 
backgrounds, because muons which decay in flight can mimic these signatures.
The authors of Ref.~\cite{Bugaev:2003sw} estimate a rate of up to 50 km$^{-3}$ 
yr$^{-1}$ of these events.  By contrast, there is no apparent physical 
background to a track which suddenly \emph{increases} in brightness as in the 
$\tau \ra \mu \nu \nu$ decay signature.  Experimental backgrounds will of 
course arise due to the intrinsic variations of lepton energy 
deposition and photon detection which contribute to the detector energy 
resolutions quoted above.  Because these variations depend on the particular
detector and analysis techniques used, Monte Carlo studies will be needed to 
quantify the background levels to be expected in actual experiments.

At lower (TeV--PeV) energies than considered here thus far, $\tau \ra \mu$ 
might also be detectable if the neutrino interaction vertex occurs within the
detector, because the two neutrinos produced in the $\tau$
decay would carry off approximately half of the lepton energy.  Although the 
$\tau$ track would not be observed directly, the presence of the tau could be
inferred on a statistical basis from the lower-than-expected energy of the 
muon track, when compared 
to the energy of the hadronic shower at the $\nu N$ vertex.  However, this 
signature would suffer from the same backgrounds as the `inverted' lollipop
(in which the tau production vertex and tau track are observed, rather than 
the tau track and decay vertex).  These signatures can be faked by a $\nu_\mu 
N$ CC interaction with high $y$, where less than the mean energy is 
transferred to the outgoing
$\mu$, or by CC $\nu_e$ or NC $\nu_x$ interactions where secondary 
lower-energy $\mu$ are produced via $\pi^\pm$ decay in the hadronic shower.  

We believe the signature of muonic tau decay will be useful in identifying 
astrophysical tau neutrino events in the coming generation of kilometer-scale
Cherenkov neutrino telescopes such as IceCube.  This signature may be 
particularly important in the energy region above a few tens of PeV, where
the classic double-bang signature is no longer observable.

\section{Acknowledgments}
The authors would like to thank John Beacom, Spencer Klein, Peter 
M\'{e}sz\'{a}ros, Irina Mocioiu, Dave Seckel, and Pat Toale for helpful 
comments and discussions, and Dima Chirkin for assistance with the software 
used to evaluate lepton energy losses.  This work was supported by NSF grants 
PHY-0244952, PHY-0554868, and AST-0307376.


\end{document}